\documentclass[preprint]{aastex631}

\shorttitle{Predicting CME Geoeffectiveness}
\shortauthors{Pricopi et al.}

\graphicspath{{./}{figures/}}

\usepackage{amsmath, tikz, makecell, graphicx, wrapfig}
\usetikzlibrary{calc} 
\graphicspath{ {./images/} }
\begin{document}

\title{Predicting the Geoeffectiveness of CMEs Using Machine Learning}

\author[0000-0001-9231-4637]{Andreea-Clara Pricopi}
\affiliation{Technical University of Cluj-Napoca,
Cluj-Napoca, Romania, 400114}

\author[0000-0002-3491-1983]{Alin Razvan Paraschiv}
\affiliation{High Altitude Observatory,
National Center for Atmospheric Research,
Boulder CO 80307-3000, USA}

\author[0000-0003-1703-8219]{Diana Besliu-Ionescu}
\affiliation{Astronomical Institute of the Romanian Academy, 
Str. Cutitul de Argint, Nr. 5,
Bucharest, Romania, 040557}

\author[0000-0001-8426-588X]{Anca-Nicoleta Marginean}
\affiliation{Technical University of Cluj-Napoca, 
Cluj-Napoca, Romania, 400114}

\begin{abstract}
Coronal mass ejections (CMEs) are the most geoeffective space weather phenomena, being associated with large geomagnetic storms, having the potential to cause disturbances to telecommunication, satellite network disruptions, power grid damages and failures.
Thus, considering these storms' potential effects on human activities, accurate forecasts of the geoeffectiveness of CMEs are paramount. This work focuses on experimenting with different machine learning methods trained on white-light coronagraph datasets of close to sun CMEs, to estimate whether such a newly erupting ejection has the potential to induce geomagnetic activity. We developed binary classification models using logistic regression, K-Nearest Neighbors, Support Vector Machines, feed forward artificial neural networks, as well as ensemble models. At this time, we limited our forecast to exclusively use solar onset parameters, to ensure extended warning times. We discuss the main challenges of this task, namely the extreme imbalance between the number of geoeffective and ineffective events in our dataset, along with their numerous similarities and the limited number of available variables. We show that even in such conditions, adequate hit rates can be achieved with these models.
\end{abstract}


\section{Introduction} 
\label{sec:intro}

The purpose of this work is to develop a machine learning (ML) based model that can predict whether a coronal mass ejection (CME) will be geoeffective, using only numerical solar parameters as input. 

Coronal mass ejections are solar eruptive events whose magnetically charged particles can, directly or indirectly, under certain circumstances, reach Earth and cause geomagnetic storms (GSs), i.e., be geoeffective. These storms represent perturbations in the Earth's magnetic field, which have the potential to lead to electrical systems and grids failure and/or damage, power outages, navigation errors, radio signal perturbations, significant exposure to dangerous radiations for astronauts during space missions, etc. Given the potential negative impacts of such storms, predicting their occurrence is paramount for enabling safeguarding of human technology
\citep{swesolar2006,  sweearth2007, NAP13060, vourlidas2019, swesolar2021}. Moreover, 
CMEs are the most geoeffective solar phenomena \citep{schwenn2005,gopalswamy2007,vourlidas2019}, being associated with more intense GSs than other, more frequent GS sources, such as high-speed streams (HSSs) and corotating interaction regions (CIRs) \citep{kilpuaetall2017}.

The intensity of the storms can be measured by various geomagnetic indices such as Ap, Kp, AE, PC or Dst \citep[see][and references therein]{lockwood2013LRSP}. Herein, we have chosen to use the values of the Dst index \citep{sugiura1964} to establish whether the magnetic field perturbations do, in fact, manifest as storms. This is an index that is calculated using four geomagnetic stations situated at low latitudes. Depending on the value of this index, it can be established whether these perturbations are associated with geomagnetic storms or not. In terms of storm intensity, one of the most popular classifications that takes into consideration the minimum value of the Dst index is that of \cite{gonzalezetal1994}. Thus, a minor storm has -30 nT $\geqslant$ Dst$_{min}$ $>$ -50 nT, a moderate storm has -50 nT $\geqslant$ Dst$_{min}$ $>$ -100 nT, while intense storms are considered to have Dst$_{min}$ $\leqslant$ -100 nT. Consequently, we only considered GSs defined by a Dst value $\leqslant$ -30 nT that were associated with ICMEs driven by CMEs.
 
Literature available prediction models can be divided by the methods used and final purpose. 
Methods based on physical processes are used to model the propagation of the CME through the interplanetary space \citep[][]{roussevlugaz07} or the interaction of two different CMEs travelling towards Earth \citep[][]{lugazetal2005, mancvo08}.

Methods based on physical processes with the purpose of predicting the arrival time of the CME at Earth are preferred for operational use, e.g., the WSA-ENLIL Cone model
being currently used by the US National Ocean and Atmospheric Administration (NOAA) for providing warnings of potential geomagnetic storms caused by Earth-directed CMEs and solar wind structures \citep[][]{Parsons2011}. 

ML methods, such as neural networks, are an emerging research area for this subject \citep[][]{vourlidas2019}. \cite{10.1093/mnras/stv2782} used a neural network and a dataset of 153 CMEs for predicting their transit times. 
\cite{wangetal2019convolution} proposed an approach based on convolutional neural networks for predicting CMEs' arrival times. They analyzed transit times from the Sun to Earth of 223 geoeffective CME events observed between 1996 to 2018 and obtained a 12.4 hours mean absolute error (MAE).  \cite{DBLP:journals/remotesensing/FuZYFLM21} used a deep learning framework with image data as inputs, for the same purpose, obtaining a MAE = 5.8 hours, as well as a 27\% F1 score and 75.1\% accuracy for the geoeffectiveness prediction.

There are non-linear logistic models that can predict the association of CMEs with intense or superintense GSs \citep[][]{srivastava2005} - where events that generated a superintense GS are considered positive, while those that generated an intense one are considered negative. \cite{srivastava2005} predicted 78\% geomagnetic storms from the training set, and 85\% from the validation one, using a dataset containing 64 CMEs observed between 1996 and 2002.
\cite{besliuionescuetal2019, bdimm2021} applied the same technique, using only solar parameters, but defined negative events as CMEs that were not associated with any GS (i.e., Dst $>$ -30 nT). 

Another type of model employed for predicting geoeffectiveness is the support vector machine (SVM). \cite{2012JKAS...45...31C} described its utility for predicting the geoeffectiveness of halo CMEs (having the angular width $>$ 120°). The dataset used for the study contained 733 events, observed between January 1996 and April 2010. The authors used the grid search method to experiment with various combinations of kernel functions and hyperparameter values, eventually obtaining 66\% accuracy, 76\% real positive rate (TPR), 49\% real negative rate (TNR), 72\% false alarm rate (FAR).

In this study, we used data covering 19 years (1996-2014) to train and test our models, taking into consideration all storms, regardless of their intensity (i.e., including minor ones). Despite evidence that interplanetary parameters influence the geoeffectiveness of CMEs \citep[e.g.][]{akasofu81,srivastava2004,yermolaevetal2005, besliuionescuetal2019}, we decided to not utilize such data, as it would limit the warning time from the order of days to that of hours, at best. Such models can then be used as early warning modules for more complex setups, involving forward modelling of solar eruptions and interplanetary propagation and interactions.

We employ a number of different supervised ML techniques for binary classification, as described in section \ref{sec:methods}, using only solar parameters as inputs. Moreover, in response to the challenges posed by the dataset (as discussed in section \ref{sec:data}), we artificially generate new virtual geoeffective CME data samples, which is an innovative aspect for this specific problem. In addition to this, we introduce the notion of uncertainty regarding the lack of geoeffective potential for the CMEs prior to a geomagnetic storm with no certain cause associated, by using sample weights. 

\section{Data and interpretation}
\label{sec:data}

\subsection{The dataset}
The dataset utilized for this research contains aggregated data from 3 different sources: the Solar and Heliospheric Observatory \citep[SOHO; ][]{domingo1995} Large Angle and Spectrometric Coronagraph \citep[LASCO;][]{lasco1995} catalog, provided by the ‘Coordinated Data Analysis Workshops (CDAW) Data Center’\footnote{\url{https://cdaw.gsfc.nasa.gov/CME\_list/,}\citep[][]{2009EM&P..104..295G}}, for the CMEs' attributes, the Solar H-alpha Flare Index (FI) \citep[][]{1987Ap&SS.135..201A} values made available by courtesy of Kandilli Solar Observatory and NOAA's 'National Geophysical Data Center (NGDC)’ \footnote{\url{https://www.ngdc.noaa.gov/stp/space-weather/solar-data/solar-features/solar-flares/index/flare-index/}}, as well as the catalog compiled by \cite{richardsoncane2010} for the correlation between the LASCO events and the Dst values of the associated geomagnetic storms, as can be seen in Table \ref{tab:dataset}. 
\begin{deluxetable}{lllllllllll}[!t]
\label{tab:dataset}
\tablehead{ \colhead{CME} & \colhead{CPA} & \colhead{AW} & \colhead{LS} & \colhead{SOSFHI} & \colhead{SOSFHF} & \colhead{SOS20RS} & \colhead{ACC} & \colhead{MPA} & \colhead{DST} & \colhead{FI}}
        \tablecaption{A sample of the aggregated dataset used for this work.}
          \startdata
            1997-01-20 09:31:00 & 281 & 72 & 175.0 & 237.0 & 115.0 & 0.0 & -3.3 & 285 & 0 & 0.470\\ 
            2014-04-01 14:00:00 & 242 & 43 & 409.0 & 468.0 & 352.0 & 125.0 & -8.3 & 237 & 0 & 1.210\\ 
            2006-07-09 20:06:00 & 54 & 19 & 142.0 & 85.0 & 199.0 & 649.0 & 17.4 & 46 & 0 & 0.000\\ 
            2010-12-21 02:48:00 & 48 & 114 & 369.0 & 300.0 & 439.0 & 449.0 & 4.6 & 62 & 0 & 0.000\\ 
            2012-05-21 09:12:00 & 208 & 41 & 645.0 & 407.0 & 882.0 & 830.0 & 22.0 & 222 & 0 & 0.560\\ 
\enddata
\tablecomments{Table 1 is published in its entirety in the machine-readable format. A portion is shown here for guidance regarding its form and content.}
\end{deluxetable} 

We primarily use a cleaned selection of CMEs in the LASCO CDAW catalog and then use the \citet{richardsoncane2010} catalog to identify geoeffective events. If a CME is not directly connected to a \citet{richardsoncane2010} Dst event, we assume it to be non-geoeffective. The implications of this are presented below.
 
Since its launch in 1995, LASCO has been producing synoptic observations with a cadence of 12 minutes per coronagraph, to the present day. Two coronagraphs, C2 and C3, observe the corona between heights of 2-6 and 3.7-32 solar radii, respectively. The catalog contains data for all the CMEs manually identified since 1996. There are 9 numerical parameters available in the catalog for each event, whose details (including computational ones) can be found in the LASCO documentation. Out of these attributes, the mass and kinetic energy have not been included in our selected dataset. This is due to the fact that the number of events for which these parameters' values were not computed amounts to approximately one third of the total number of CMEs in the catalog. As such, the information provided by these two parameters proved to be insufficient for this research. All other attributes available in the catalog (i.e., Central Position Angle - CPA, Angular Width - AW, Measurement Position Angle - MPA, Linear Speed - LS, 2$^{nd}$ order Speed at final height, 2$^{nd}$ order Speed at 20 Rs, Acceleration - Acc) were taken into consideration. In order to obtain a "clean" dataset, all rows from the catalog (i.e., events) containing missing values in any of the columns (e.g. 2000/03/05, 12:54:05) have been removed. This amounts to 461 events, out of which 108 miss all speed-related values, together with Acc, and 353 have non-null LSs, but missing the other 3 parameters' values. Analyzing the LASCO catalog, it can be noted that CMEs missing linear speeds were considered very poor events, unable to be properly measured. After accounting for all discussed above, we report one geoeffective event (2005/05/13 17:12, Dst = -247 nT) was lost as a result of the dataset cleaning. 

\begin{figure}[!h]
    \centering
    \begin{minipage}[b]{0.45\textwidth}
        \includegraphics[height=0.23\textheight]{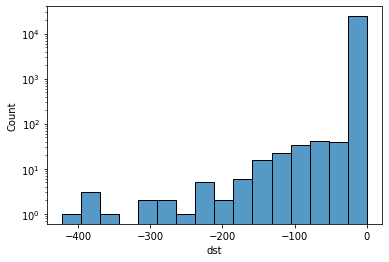}
    \end{minipage}
    \begin{minipage}[b]{0.45\textwidth}
        \includegraphics[height=0.23\textheight]{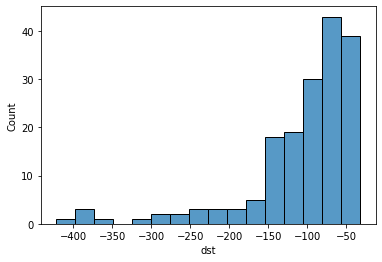}
    \end{minipage}
    \caption{Left: Histogram (logscale) of all Dst values in the dataset. Right: Histogram of the Dst values $\leqslant$ -30}
    \label{fig:histogram}
\end{figure}

Since predicting the geoeffectiveness is fundamentally a binary classification task, an output label is necessary. Therefore, we defined the labels as following:
\begin{itemize}
    \setlength\itemsep{-0.5em}
    \item 1 (i.e., positive event; geoeffective CME) for CMEs associated with Dst $\leqslant$ -30 nT;
    \item 0 (i.e., negative event; CME that was not geoeffective) for CMEs associated with Dst $>$ -30 nT;
\end{itemize}
The events labeled 1 are computed using the \cite{richardsoncane2010}, based on their association with a Dst value $\leqslant$ -30 nT (172 out of 24403 events) where an ICME had an associated CDAW event. All other CMEs, either not present in the catalog or correlated with a Dst value $>$ - 30 nT, were labeled 0 (24231 out of 24403 events). The distribution of the Dst values in the dataset can be observed in Figure \ref{fig:histogram} left and right.

In addition to the above-mentioned parameters, we use the FI as input for our models, similar to  \cite{uwamahoro2012}'s approach concerning the use of neural networks for predicting the geoeffectiveness of halo CMEs. The index's purpose is to quantify the daily flare activity. We have utilized the FI based on findings of \citet[][]{2004} where the association of some CMEs (i.e., fast, full-halo events originating from low and middle latitudes, close to central meridian) with flares is a significant driver for intense geomagnetic storms. Importantly, the FI does not address individual events, but daily averages of flaring activity. Thus, this parameter does not represent a one to one flare to CME association.

On this topic, we also mention the study of \citet{yashiroetal2006} that focused on the individual associations between CMEs and flares. The authors provide a maintained list that is available online\footnote{\url{https://cdaw.gsfc.nasa.gov/pub/yashiro/flare_cme/fclist_pub.txt}}. We have cross-correlated this list with our subset of geoeffective CME's and found that $<$40\% of combined CDAW and \citet{richardsoncane2010} events could be associated with a flare. Due to the extreme class imbalance issues discussed below in sec. \ref{sec:challenges}, we considered that removing more than half of our already small geoeffective subset would result in weak model performances. A manual association attempt on the remaining events could not be confidently established by us. Thus, we did not utilize any individual CME to flare association throughout this work.

Due to the FI only being made available for the 1976-2014 timeframe, we needed to also restrict the data from the LASCO catalog up until 2014 only. Therefore, the collected data covers the period between 1996 and 2014. After the removal of samples containing empty cells, as mentioned above, and restricting the period from 1996 to 2014, our final dataset contains 24403 CMEs, out of which 172 are geoeffective.

Using the de facto approach, we manually selected the independent features to be used as inputs. Therefore, the features used for all models were CPA, AW, LS, Acc, FI, unless specified otherwise. However, we also experimented with an automatic feature selection by reducing the total number of features using Principal Component Analysis \citep[PCA,][]{jolliffe2005principal}. PCA works by redefining data attributes by selecting the first k components in descending order of their variances (average of the squared differences from the mean). The feature redefinition refers to creating the principal components, which are linear combinations of the initial variables. From a visual perspective, the principal components can be interpreted as the directions associated with the biggest variances (i.e., most dispersed values) in a multidimensional scatterplot. These directions are, in fact, eigenvectors, while the eigenvalues signify how important the directions are, or, in other words, how much of the most significant information regarding any data point they represent. Choosing the highest variance values when lowering the number of attributes is, therefore, important in order to keep as much information from the original data as possible.
The PCA implementation utilized (see section \ref{sec:methods}) has the option of automatically choosing the number of principal components by using Minka's Maximum Likelihood Estimates method \citep[MLE,][]{Minka2000AutomaticCO}. Using this, we obtained a dataset with only one less feature. Therefore, we also chose to use k = 5 (the number of independent variables in the dataset) for the number of components to be returned. It is important to note that the new features obtained as a result of using this method are harder to interpret. 

In order to improve both the models' performances and the training times, a feature scaling preprocessing step was used. Both standard scaling  (eq. \ref{eq:std_scaling}) and l2 normalization (i.e., scaling samples to have unit norm) were tested. However, standard scaling led to the best results in all cases.
\begin{equation}
    \label{eq:std_scaling}
    z = \frac{x - \mu}{\sigma}, 
    \textnormal{where $\mu$ = mean of training samples and $\sigma$ = standard deviation}\
\end{equation}

\subsection{Challenges in data interpretation}\label{sec:challenges}
One of the most salient observations regarding the data, which is also the main obstacle for creating high-performance, reliable models (i.e. binary classifiers in this case) is the extreme class imbalance, as only 0.71\% of the events in the dataset are geoeffective. ML models are sensitive to skewed distributions \citep[][]{2016, article, inbook}, resulting in a tendency to more frequently predict the label of the majority class. In consequence, models will commonly appear to yield high proportions of correctly labeled samples for imbalanced classification problems. In other words, they will have high accuracy values. However, this would be a misleadingly optimistic performance evaluation, since this could translate into simply outputting the label with the most occurrences, instead of actually distinguishing between classes. This discussion is extended in section \ref{sec:methods}. 
In this particular case, a dummy, untrained predictor outputting only the label of the majority class would have 99.29\% accuracy, corresponding to the proportion of 99.29\% of events in the dataset that were not geoeffective. 
However, the model would identify 0 positive events, failing to achieve its primary goal of forecasting the geoeffective CMEs, despite the high accuracy value. We include the accuracy metrics for methodological completeness reasons, but we stress out that this metric in particular is of less importance when interpreting model performance.

Therefore, since the minority class is our main focus, due to the potential negative effects of the events belonging to it, the imbalance issue needs to be addressed. 
Our attempts at overcoming this challenge include using class weights \citep[][]{krawczyk2016learning} and creating artificial data samples for the minority class using the Synthetic Minority Oversampling TEchnique \citep[SMOTE,][]{2002, 2018art}.

In order to better understand the necessity for class weights, a few elementary notes on the principles behind the models described in this paper are required. A model learns to map inputs to outputs (i.e., to predict/output values close to the real ones), through training. Generally, training refers to making adjustments to several parameters, according to the error between the real output and the predicted one, in order to minimize it.

Given that, as previously stated, the geoeffective CMEs are of more concern, information about their importance needs to be embedded in the model, which can be achieved with the help of weights. The weights are used together with the error between the real output and the predicted one for determining the next adjustments to the parameters. For this specific case, the desired outcome is a more significant change of the model's parameters as a consequence of the model predicting 0 instead of 1. In other words, a model should be penalized more severely when misclassifying a more important (here, geoeffective) event.  

For this research, we considered the importance of an event to be the weight of the class it belongs to ($weight_{c}$ for class c), which is inversely proportional to the frequency of the class' appearances in the dataset (eq. \ref{eq:weight}).
\begin{equation}
    \label{eq:weight}
    weight_c = \frac{\textnormal{total number of samples}}{\textnormal{number of classes * number of samples belonging to class c}}
\end{equation}

Thus, the weights associated with the samples are 69.72 for the events belonging to the minority class and 0.5 for those in the majority class. 

The other technique for overcoming the class imbalance is oversampling. This refers to the creation of additional samples, most commonly belonging to an underrepresented class. One way of achieving this is \emph{random oversampling}, meaning creating multiple copies of some of the samples from the minority class, selected at random. However, this is similar to using higher weights for those samples, by feeding them more times to the model. Therefore, we chose to use the Synthetic Minority Oversampling Technique (SMOTE) to create new samples altogether, as described in section \ref{sec:class_imbalance}.  

Another observed issue is the high similarity degree between samples from different classes, known as \emph{class overlap}. Visually, this problem manifests as two or more points from different classes overlaying each other. This can be visualized in the figures below, representing only the negative events (fig. \ref{fig:UMAP} left), as well as both negative and positive ones (fig. \ref{fig:UMAP} right) in the dataset, respectively. In Figure \ref{fig:UMAP} right, it can be observed that there are points belonging to the positive class overlapping negative ones. In other words, some samples have approximately equal probabilities of belonging to different classes (in this case 0 and 1), turning this prediction into a particularly challenging task. 
\begin{figure}[!h]
    \centering
    \begin{minipage}[b]{0.45\textwidth}
        \centering
        \includegraphics[height=0.17\textheight]{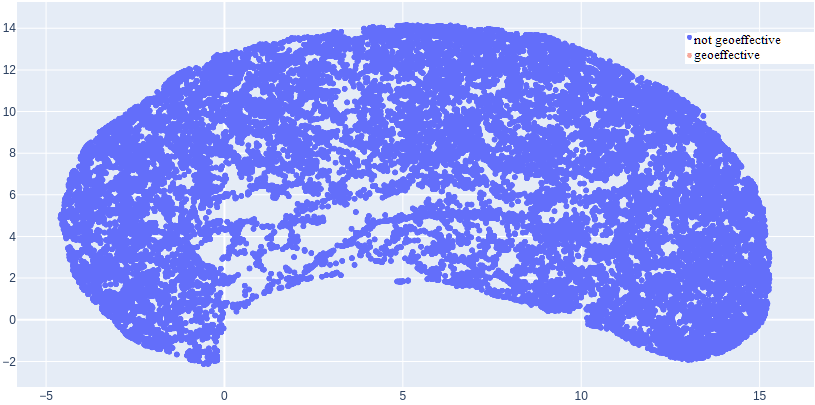}
    \end{minipage}
    \begin{minipage}[b]{0.45\textwidth}
        \centering
        \includegraphics[height=0.17\textheight]{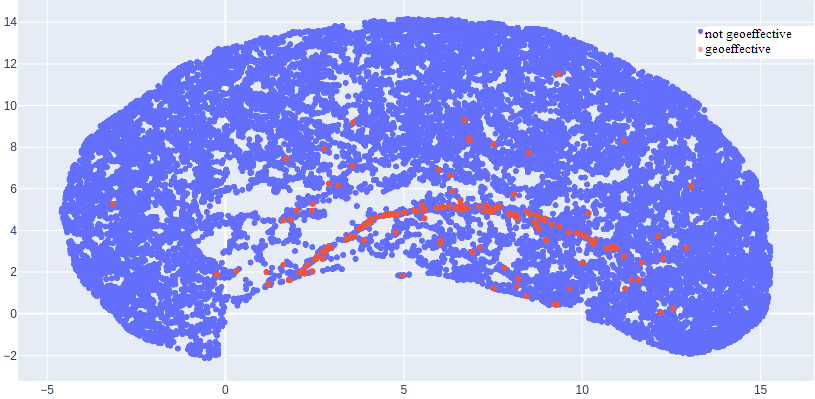}
    \end{minipage}
    \caption{Left: UMAP of the negative events only. Right: UMAP of both negative and positive events}
    \label{fig:UMAP}
\end{figure}
The 2D plots presented in fig \ref{fig:UMAP} left and rigth were obtained with the help of the dimensionality reduction technique called Uniform Manifold Approximation and Projection \citep[UMAP,][]{mcinnes2020umap}. Dimensionality reduction refers to data transformations meant to reduce the number of attributes used to describe the essential information embedded in the data. The use of this technique enabled the 2D display of the data, which provides visual insight into the it. While, arguably, the plots are oversimplified, overlapping is apparent by data analysis, which shows substantial similarities in the feature values of the two classes. \cite{VUTTIPITTAYAMONGKOL2021106631} show that class overlap is a highly overlooked issue which negatively impacts classifications, amplifying the negative effects of class imbalance. Due to the overlap, the models tend to misclassify positive instances in the proximity of the class boundaries, which will be discussed in section \ref{sec:conclusion}.

We hypothesize that one of the causes behind the class overlap issue is the limited number of independent features used for distinguishing between positive and negative events. We consider the location of the origin of the CME and its direction \citep[][]{2005} as examples of such features, that are not adequately contained in any of the catalogs we used. 
Additionally, information that is also likely to reduce the class overlap is the CMEs' association with flares, as studied by \cite{yashiroetal2006}.   

Another potential issue is the presumed incompleteness of the known associations between geomagnetic storms and the causal CMEs identified by \citet{richardsoncane2010}. In other words, not all the disturbances identified by \citet{richardsoncane2010} are correlated with a specific LASCO identified event, which allows for presumptions regarding their cause. One reason behind this issue is the fact that, due to the variance of travel times, it is not always possible to determine which, out of several different CMEs, was the unquestionable cause of a specific disturbance. This means that, given that we consider all CMEs not found in the catalog compiled by \cite{richardsoncane2010} to not be geoeffective, we could be training the models with inaccurate information (i.e. events labeled as 0 due to lack of certainty regarding their exact cause that are, in fact, geoeffective).
In the \cite{richardsoncane2010} catalog, there are 227 events with no associated LASCO CME cause having occurred until 2015, out of which 131 have Dst $\leqslant$ -30.  
Following the approach described by \cite{uwamahoro2012}, we consider all CMEs from 15 hours and up to 5 days prior to any of the 131 geoeffective disturbances identified by \cite{richardsoncane2010} before 2015 that have no association with a LASCO event to be \emph{possibly} geoeffective. This means that neither the label 0, nor 1 is unequivocal for these events.

We experimented with capturing such information about potentially geoeffective events by using example weights. For this investigation, the samples were not assigned weights based only on the class they belonged to, as previously described. All potentially geoeffective events, as defined above, were assigned half the weight of an event that was not geoeffective. This way, if the model predicted a potentially geoeffective event as geoeffective, it would not be penalized as highly as it would be for misclassifying an event that has an unambiguous label.

\section{Machine Learning methods, their applications and results}
\label{sec:methods}
For this research, we studied and compared the following binary classifiers from the perspective of predicting CMEs' geoeffectiveness: logistic regression, K-Nearest Neighbors, Support Vector Machines, feed forward artificial neural networks and ensemble models. For the implementations of the various models and algorithms (e.g., preprocessing, grid search, cross validation, PCA), the scikit-learn library \citep[][]{scikit-learn} was used, unless stated otherwise. In addition to this, the artificial neural networks were created with the help of TensorFlow 2 \citep[][]{tensorflow2015-whitepaper} and Keras \citep[][]{chollet2015keras}.

In this section, the principled inner-workings of the models employed will be described, together with our actual experiments and the hyperparameter options we explored. Hyperparameters are model parameters that are not learned and need to be submitted beforehand. Finding their optimal values is one of the most important aspects of creating a model and represents a process called hyperparameter tuning, which generally involves testing various inputs. For this study, with the exception of the neural networks, this process was performed using the grid search approach. Grid search involves specifying the values to be tested for each desired hyperparameter, exhaustively trying all the possible combinations and comparing the yielded performances in order to find the optimal set of values.  
The significance of each hyperparameter we tuned, together with the values used, will be detailed for each model. 

When evaluating the performance of our models, we analyzed and compared a number of performance metrics (recall, precision, F1 score, accuracy and specificity), as detailed in the following paragraphs. For a better understanding of the terms used for defining the metrics and their evaluation, the essential notations used throughout this paper are the following:
\begin{itemize}
    \setlength\itemsep{-0.5em}
    \item TP - True Positive (event correctly predicted to be geoffective)
    \item TN - True Negative (event correctly predicted as not geoeffective)
    \item FP - False Positive (false alarm)
    \item FN - False Negative (undetected geoefective event)
\end{itemize}    

We considered the recall (eq. \ref{eq:recall}) - also known as sensitivity - to be the main performance indicator, given the high cost associated to not anticipating a potential storm (e.g., without any warning, electrical/satellite/communication systems would not be safeguarded). The recall value should be interpreted as the percentage of the known geoeffective events that were correctly identified by the model.
\begin{equation}
    \label{eq:recall}
    recall = \frac{TP}{TP + FN}
\end{equation}

Recall's complementary performance metric is the precision (eq. \ref{eq:precision}), which indicates how many of the events predicted to be geoeffective \emph{actually} were geoeffective.
\begin{equation}
    \label{eq:precision}
    precision = \frac{TP}{TP + FP} = 1 - \textnormal{False Alarm Ratio}
\end{equation}

Despite the fact that the recall and precision can each be improved at the expense of the other, it is decidedly desirable for both their values to be as high as possible. A good indicator of their balance is given by the F1 score (eq. \ref{eq:f1score}):
\begin{equation}
    \label{eq:f1score}
    F1 score = \frac{2 * (precision * recall)}{precision + recall}
\end{equation}

Additionally, we also computed the accuracy (eq. \ref{eq:accuracy}), which is equal to the proportion of correct predictions. We reiterate that the accuracy is not a reliable metric by itself in the case of binary classification on imbalanced data.

\begin{equation}
    \label{eq:accuracy}
    accuracy = \frac{TP + TN}{TP + TN + FP + FN}
\end{equation}

Finally, the specificity (eq. \ref{eq:specificity}) refers to the percentage of correctly labeled negative events.
\begin{equation}
    \label{eq:specificity}
    specificity = \frac{TN}{TN + FP}
\end{equation}

For testing the performances of our models, we used stratified k-fold cross validation \citep[][]{Refaeilzadeh2016}, with the default value k = 5. This means the data is shuffled and split into 5 subsets, each preserving the approximate proportion of positive to negative samples in the original dataset. The models are trained on 4 out of 5 subsets and tested on the remaining one. The process is repeated until the model has been trained and tested on all subsets (i.e., 5 times). The performance values are the average for all tests. Therefore, this evaluation method is expected to yield less biased results, given that the models are not tested only on a fraction of the data, which could have various particularities, making it easier or harder to predict on. However, in order to obtain a visual representation of the predictions with the help of UMAP (similar to those in Figure \ref{fig:UMAP} left and right, a single test set had to be chosen. For this, we used 20\% of the data, split in a stratified manner, as further described in section \ref{sec:ensemble}.
In the following subsections, the methods and models used for the experiments are detailed.

\subsection{Logistic Regression}
\label{sec:lin_reg}
The logistic regression is a statistical model used to obtain a dependent variable (the label) based on one or more independent ones (the attributes). The predicted value can be interpreted as the probability
of the event belonging to the positive class (e.g.: a value of 0.2 represents a 20\% chance that the real label is 1, leading to the choice of the predicted label to be 0). This is expressed in eq. \ref{eq:probability}, where X represents all the examples, while y is one real target value and $\hat{y}$ is a predicted one:
\begin{equation} \label{eq:probability}
    \hat{y} = P(y = 1|X), \;\;\; where
\end{equation}
\begin{equation} \label{eq:logReg}
    \hat{y} = \sigma(W^{T}X + b), \;\;\; and
\end{equation}
\begin{equation} \label{eq:sigma}
    \sigma(z) = \frac{1}{1 + e^{-z}}
\end{equation}

Generally, in the case of normal distributions, the threshold used for discriminating between the 2 classes is 0.5, but, in practice, the threshold may depend on the context of the problem and, implicitly, the class imbalance \citep[][]{2018book}. The adjustments of the threshold is known as \emph{threshold moving}.

Since the distribution of the classes in the dataset is not normal, we expected that the separation between the classes using the threshold associated to normal distributions to be susceptible to improvement. Given that there is no a priori knowledge of a cost associated with missclassifications, we have tested values between 0.5 and 0.9 for the threshold, through cross validation. Similarly, \cite{DBLP:journals/remotesensing/FuZYFLM21} analyzed the influence of 20 different thresholds between 0 and 1, observing performance improvements when choosing a value higher than 0.5.

We discovered empirically that for this model, the threshold leading to the highest F1 score (therefore the best balance between recall and precision) is 0.9. The performances obtained with all other tested thresholds are summarized in Table \ref{tab:linreg_thresholds}. 

\begin{table}[h]
    \centering
    \caption{Performances of the linear regression model with class-balanced weights with various thresholds, on the entire dataset, after 5-fold cross-validation}
    \begin{tabular*}{\textwidth}{l@{\extracolsep{\fill}}ccccc}
        \hline
         LR Model Threshold &  0.5 & 0.6 & 0.7 & 0.8 & 0.9\\
         \hline
         Recall &  89.24\% & 86.63\% & 82.00\% & 79.21\% & 73.24\%\\
         Precision &  9.79\% & 11.39\% & 12.98\% & 14.59\% & 16.29\%\\
         F1 score &  17.64\% & 20.13\% & 22.39\% & 24.56\% & 26.63\%\\
         Accuracy &  94.11\% & 95.13\% & 95.96\% & 96.61\% & 97.13\%\\
         Specificity &  94.13\% & 95.14\% & 96.05\% & 96.73\% & 97.28\%\\
         \hline
    \end{tabular*}
    \label{tab:linreg_thresholds}
    \vspace{-4mm}
\end{table}
\begin{table}[h!]
\centering
\setlength{\tabcolsep}{-0.25pt}
    \caption{Performances of linear regression models with threshold = 0.9, on the entire dataset, after 5-fold cross-validation}
    \begin{tabular*}{\textwidth}{l@{\extracolsep{\fill}}cccc}
            \hline
             LR Model & Balanced weights & Uncertainty-based weights & With PCA, 5 components\\
             \hline
             Recall & 73.24\% & 72.65\% & 72.65\%\\
             Precision & 16.29\% & 15.98\% & 16.60\%\\
             F1 score & 26.63\% & 26.18\% & 27.02\%\\
             Accuracy & 97.13\% & 97.11\% & 97.23\%\\
             Specificity & 97.28\% & 97.28\% &  97.40\%\\
             \hline
    \end{tabular*}
    \label{tab:linreg_cv}
    \vspace{-4mm}
\end{table}

Using example weights based on our certainty regarding whether the events were geoeffective or not did not lead to remarkable performance changes, while training the model using 5 principal components instead of the manually selected independent features led to precision increase, as shown in Table \ref{tab:linreg_cv}.

\subsection{K-Nearest Neighbors}
\label{sec:knn}
K-Nearest Neighbors (KNN) is a similarity-based algorithm, where similarity is expressed in terms of a generalized concept of distance. Various ways of computing the distance can be used, e.g., Manhattan (eq. \ref{eq:manhattan}), euclidean (eq. \ref{eq:euclidean}), Minkowski (eq. \ref{eq:minkowski}). KNN models assign the label of an event based on the label of the majority of the k most similar examples, where k is predefined (not to be confused with the number of folds used for cross-validation, which is an independent variable). This is done by following the steps below:

\begin{enumerate}
    \setlength\itemsep{-0.5em}
    \item Compute the distance between the point (example) whose label needs to be predicted and all the labeled points in the dataset.
    \item Choose the first k smallest distances (i.e., the k-nearest neighbors).
    \item Assign the label of the majority of the k-nearest neighbors to the unlabeled point.
\end{enumerate}

Simply using the predominant label in a group of k neighbors to decide upon the label of a new sample is known as using uniform weights. However, for extremely imbalanced datasets, such as the one used for this research, there is a high chance the majority of a sample's neighbors belong to the majority class. This means that regardless of how similar a sample might be to another one belonging to the minority class, the low number of such examples might not be enough to form a majority in many cases, resulting in numerous misclassifications. One way of tackling this issue is by using distance-based weights when determining the label. In other words, the smaller the distance between a sample and its neighbor (i.e. the more similar they are), the more the label of the latter will weigh when deciding what the prediction should be. Therefore, the role of distance-based weights is to ensure that minority samples have the potential to counterbalance the high number of majority samples.
For this study, we have experimented with various values for k, in order to find an optimal one, bearing in mind that too small a value could make the model sensitive to noise, while too big a value eventually leads to no performance improvements, at the cost of computation time. At the same time, we have also attempted to use different weights and distances.
The values tested using grid search were:

\begin{itemize}
    \setlength\itemsep{-0.5em}
    \item k = 3, 5, 7;
    \item weight: uniform and distance-based;
    \item distance: Manhattan (eq. \ref{eq:manhattan}), euclidean (eq. \ref{eq:euclidean}), Minkowski (eq. \ref{eq:minkowski}) with p = 3;
\end{itemize}
\begin{equation}
    \label{eq:manhattan}
    d(x, y) = \sum{|x - y|}
\end{equation}
\begin{equation}
    \label{eq:euclidean}
    d(x, y) = \sqrt{\sum{(x - y)^{2}}} 
\end{equation}
\begin{equation}
    \label{eq:minkowski}
    d(x, y, p) = (\sum{|x- y|^{p}})^{\frac{1}{p}}
\end{equation}

The performances of the best KNN models are summarized in Table \ref{tab:knn_cv}. Our grid-search experiments showed the optimal number of neighbors in this case to be 7 when using uniform weights and 5 when using distance-based weights. The KNN models attained the highest precision values out of all tested models, while, in turn, also having the lowest recall values. The best precision values were obtained using uniform weights and the Minkowski distance, with p = 3. Simply using a distance-base weighted approach, with the same hyperparameters, led to no recall improvement, accompanied by an $\sim$10\% precision decrease, which could indicate the need for better signaling great differences between the values of various examples' features. As indicated by the grid search, Manhattan distance is a more suitable distance metric. Nonetheless, the discrepancy between the recall increase and the precision decrease when using distance-based weights, as opposed to uniform weights, emphasizes the models' difficulty in overcoming the class imbalance and class overlap. In terms of similarity, additional information could have a great impact on distinguishing between the events that are geoeffective and those that are not. 

\begin{table}[!ht]
\centering
    \caption{Performances of KNN models on the entire dataset, after 5-fold cross-validation}
    \begin{tabular}{lccc}
            \hline
             KNN Model&\makecell[l]{\\7 neighbours\\Minkowski distance\\uniform weights} & \makecell[l]{\\5 neighbours\\Manhattan distance\\distance-based weights} & \makecell[l]{PCA, 5 components\\ 7 neighbours \\Minkowski distance\\ uniform weights} \\
             \hline
             Recall &  4.03\% & 7.51\% & 2.89\%\\
             Precision &  67.99\% & 22.94\% &  60.00\%\\
             F1 score &  7.49\% & 11.17\% & 5.49\%\\
             Accuracy &  99.29\% & 99.19\% & 99.30\%\\
             Specificity & 99.97\% & 99.85\% & 99.98\%\\
             \hline
    \end{tabular}
    \label{tab:knn_cv}
    \vspace{-4mm}
\end{table}

After 5-fold stratified cross-validated predictions, the KNN model with uniform weights correctly identified 3 geoeffective CMEs, all being full halo events (namely, the CMEs from 2000-07-14 10:54, 2002-08-16 12:30, 2004-07-25 14:54). 

The model also predicted 12 false positives, all having the CPA and AW equal to 360 (i.e., full halos). An example of such an event is the CME from 2012-03-07 01:30. The false alarm's attributes are presented in Table \ref{tab:neighbors}, together with those of its nearest neighbors, listed from closest to farthest. It can be observed that, out of the 7 events considered to be most similar to it, 5 were geoeffective, which, consequently, led to the model predicting the event to be geoeffective as well. The model's high number of false positives, compared to its number of true positives, is due to the high number of geoffective CMEs having AW = 360 in our dataset (113 out of 172), in addition to the even higher number of CMEs having AW = 360 that were not geoeffective (537) and suggests that there is additional information needed to better differentiate between the full halo CMEs that are geoeffective and those that are not.

\begin{table}[!ht]
\centering
    \caption{Example of a false alarm detected by the KNN model with uniform weights, together with its nearest neighbors}
    \begin{tabular*}{\textwidth}{c@{\extracolsep{\fill}}ccccc}
            \hline
             CME & LS & Acc & FI & Dst & Geoeffective\\
             \hline
             2012-03-07 01:30 & 1825 & -160.9 & 36.29 & 0 & 0\\
             \hline
             2003-10-29 20:54 & 2029 & -146.5 & 31.88 & -383 & 1\\
             2005-09-10 21:52 & 1893 & -171.7 & 43.74 & 0 & 0\\
             2005-01-17 09:30 & 2094 & -118.8 & 33.60 & -80 & 1\\
             2000-07-14 10:54 & 1674 & -96.1 & 25.15 & -301 & 1\\ 
             2002-08-16 12:30 & 1585 & -67.1 & 29.66 & -106 & 1\\
             2003-11-02 09:30 & 2036 & -64.2 & 30.47 & 0 & 0\\
             2006-12-13 02:54 & 1774 & -61.4 & 47.55 & -162 & 1\\
             \hline
    \end{tabular*}
    \label{tab:neighbors}
    \vspace{-4mm}
\end{table}

\subsection{Support Vector Machines}

The Support Vector Machines \citep[SVM,][]{1995} aim to find the optimal hyperplane delimiting the points belonging to 2 classes (while the method can actually be used for multiple classes, this is outside the scope of this work). 
The hyperplane represents an ensemble of points in an n-1 dimensional vector space, where n is the number of attributes used (e.g., for bidimensional datasets, i.e., having only 2 attributes, the hyperplane, is, in fact, a straight line; for 3D data, the classes will be separated by a 2D plane, etc). The essence of this algorithm lies in the way the optimal hyperplane is found, which is by ensuring that the area between the plane and the closest points (which are called the support vectors) is maximal.
In the case of binary classification with linearly separable classes, the decision function is:
\begin{equation} \label{eq:svmLine}
    W^{T}X + b = 0
\end{equation}

Thus, the predicted labels are as follows:
\begin{equation} \label{eq:svmPredictions}
    \hat{y} = \begin{cases}
   0, & \text{if W$^{T}$X + b $<$ 0}\\
   1, & \text{if W$^{T}$X + b $\geq$ 0}
    \end{cases}
\end{equation}

Ensuring that the separation area is maximal is equivalent to finding the W that maximizes the value of \(\frac{2}{||W||}\).
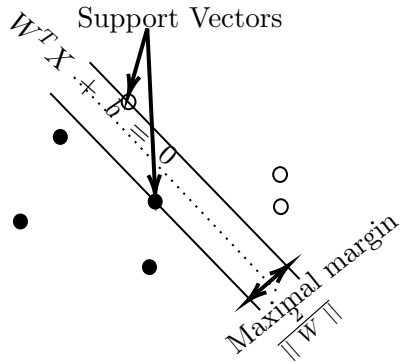
\begin{figure}
    \centering
        \caption{Illustration of the principles behind the SVM approach}
        \label{fig:SVM}
        \tikzset{every picture/.style={line width=0.75pt}} 
\begin{tikzpicture}[x=0.75pt,y=0.75pt,yscale=-0.7,xscale=0.7, baseline=(XXXX.south) ]
\path (0,284);\path (337,0);\draw    ($(current bounding box.center)+(0,0.3em)$) node [anchor=south] (XXXX) {};
\draw    (35.32,66.76) -- (186.58,223.17) ;
\draw    (63.75,44.51) -- (92.43,74.16) -- (215.02,200.91) ;
\draw  [fill={rgb, 255:red, 0; green, 0; blue, 0 }  ,fill opacity=1 ] (9.16,159.3) .. controls (9.16,156.52) and (11.26,154.27) .. (13.85,154.27) .. controls (16.45,154.27) and (18.55,156.52) .. (18.55,159.3) .. controls (18.55,162.07) and (16.45,164.33) .. (13.85,164.33) .. controls (11.26,164.33) and (9.16,162.07) .. (9.16,159.3) -- cycle ;
\draw  [fill={rgb, 255:red, 0; green, 0; blue, 0 }  ,fill opacity=1 ] (37.88,98.32) .. controls (37.88,95.54) and (39.98,93.29) .. (42.57,93.29) .. controls (45.16,93.29) and (47.26,95.54) .. (47.26,98.32) .. controls (47.26,101.1) and (45.16,103.35) .. (42.57,103.35) .. controls (39.98,103.35) and (37.88,101.1) .. (37.88,98.32) -- cycle ;
\draw  [fill={rgb, 255:red, 0; green, 0; blue, 0 }  ,fill opacity=1 ] (102.14,192.22) .. controls (102.14,189.45) and (104.24,187.19) .. (106.83,187.19) .. controls (109.42,187.19) and (111.52,189.45) .. (111.52,192.22) .. controls (111.52,195) and (109.42,197.25) .. (106.83,197.25) .. controls (104.24,197.25) and (102.14,195) .. (102.14,192.22) -- cycle ;
\draw  [fill={rgb, 255:red, 0; green, 0; blue, 0 }  ,fill opacity=1 ] (106.26,144.97) .. controls (106.26,142.19) and (108.36,139.94) .. (110.95,139.94) .. controls (113.54,139.94) and (115.64,142.19) .. (115.64,144.97) .. controls (115.64,147.74) and (113.54,150) .. (110.95,150) .. controls (108.36,150) and (106.26,147.74) .. (106.26,144.97) -- cycle ;
\draw   (86.79,73.01) .. controls (86.79,70.07) and (89.01,67.68) .. (91.76,67.68) .. controls (94.51,67.68) and (96.74,70.07) .. (96.74,73.01) .. controls (96.74,75.96) and (94.51,78.35) .. (91.76,78.35) .. controls (89.01,78.35) and (86.79,75.96) .. (86.79,73.01) -- cycle ;
\draw   (196.54,148.63) .. controls (196.54,145.68) and (198.76,143.29) .. (201.51,143.29) .. controls (204.26,143.29) and (206.49,145.68) .. (206.49,148.63) .. controls (206.49,151.57) and (204.26,153.96) .. (201.51,153.96) .. controls (198.76,153.96) and (196.54,151.57) .. (196.54,148.63) -- cycle ;
\draw   (195.97,125.45) .. controls (195.97,122.51) and (198.2,120.12) .. (200.94,120.12) .. controls (203.69,120.12) and (205.92,122.51) .. (205.92,125.45) .. controls (205.92,128.4) and (203.69,130.79) .. (200.94,130.79) .. controls (198.2,130.79) and (195.97,128.4) .. (195.97,125.45) -- cycle ;
\draw [line width=1.5]    (105.27,20.11) -- (92.58,64.79) ;
\draw [shift={(91.76,67.68)}, rotate = 285.85] [color={rgb, 255:red, 0; green, 0; blue, 0 }  ][line width=1.5]    (14.21,-4.28) .. controls (9.04,-1.82) and (4.3,-0.39) .. (0,0) .. controls (4.3,0.39) and (9.04,1.82) .. (14.21,4.28)   ;
\draw [line width=1.5]    (105.27,20.11) -- (110.81,136.94) ;
\draw [shift={(110.95,139.94)}, rotate = 267.28] [color={rgb, 255:red, 0; green, 0; blue, 0 }  ][line width=1.5]    (14.21,-4.28) .. controls (9.04,-1.82) and (4.3,-0.39) .. (0,0) .. controls (4.3,0.39) and (9.04,1.82) .. (14.21,4.28)   ;
\draw  [dash pattern={on 0.84pt off 2.51pt}]  (52.38,55.18) -- (199.66,210.97) ;
\draw [line width=1.5]    (179.75,214.5) -- (204.79,192.82) ;
\draw [shift={(207.06,190.85)}, rotate = 499.1] [color={rgb, 255:red, 0; green, 0; blue, 0 }  ][line width=1.5]    (14.21,-4.28) .. controls (9.04,-1.82) and (4.3,-0.39) .. (0,0) .. controls (4.3,0.39) and (9.04,1.82) .. (14.21,4.28)   ;
\draw [shift={(177.49,216.46)}, rotate = 319.1] [color={rgb, 255:red, 0; green, 0; blue, 0 }  ][line width=1.5]    (14.21,-4.28) .. controls (9.04,-1.82) and (4.3,-0.39) .. (0,0) .. controls (4.3,0.39) and (9.04,1.82) .. (14.21,4.28)   ;
\draw (185.99,240.43) node [anchor=north west][inner sep=0.75pt]  [rotate=-318.57]  {$\frac{2}{||\ W\ ||}$};
\draw (52.76,2.8) node [anchor=north west][inner sep=0.75pt]   [align=left] {Support Vectors};
\draw (157.55,243.7) node [anchor=north west][inner sep=0.75pt]  [rotate=-321.23] [align=left] {Maximal margin};
\draw (15.7,0.95) node [anchor=north west][inner sep=0.75pt]  [rotate=-41.49]  {$W^{T} X\ +\ b\ =\ 0$};
\end{tikzpicture}
\end{figure}        

\begin{figure}[!ht]
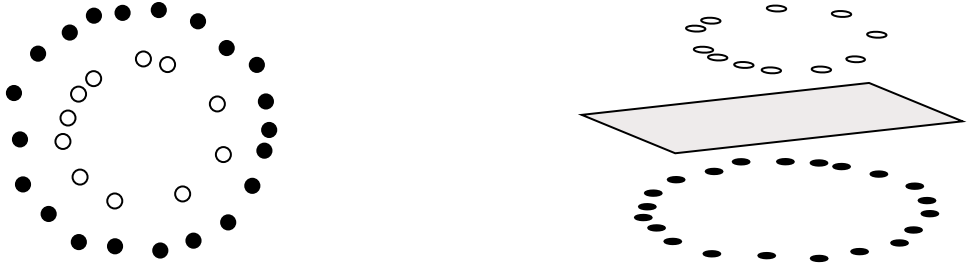

    \centering
    \begin{minipage}{0.45\textwidth}
        \centering
        \input figures/nonseparabletikz.tex
    \end{minipage}
    \begin{minipage}{0.45\textwidth}
        \centering
        \input figures/separabletikz.tex
    \end{minipage}
    \caption{Left: data that is not linearly separable. Right: hyperplane separating data in a 3D space}
    \label{fig:SVM2}
\end{figure}

However, not all datasets are initially linearly separable (e.g., Figure \ref{fig:SVM} and Figure \ref{fig:SVM2} left). Nevertheless, it is generally considered that there does exist a higher dimensional feature space in which the data can be linearly separated \citep[Cover's Theorem,][]{4038449, haykin2009neural} (fig. \ref{fig:SVM2} right). Therefore, mapping the initial values to this new space is required.

Since this transformation can be computationally demanding, the \emph{kernel trick} \citep[][]{1000150} is used. The kernel is a function whose inputs are vectors (in the original space) returning their dot product in the feature space, instead of explicitly computing the new coordinates. Thus, a kernel function k is defined as:

\begin{equation} \label{eq:kernel}
    k(a, b) = \langle{\phi(a), \phi(b)}\rangle
\end{equation}
for a, b \(\in\) X, where \(\phi : X \rightarrow \Re^n\).

For SVMs, the kernel function is a hyperparameter itself. The ones we have experimented with are: linear (eq. \ref{eq:linKernel}, also known as a "non-kernel", since it does not project the data onto a higher dimension), polynomial (eq. \ref{eq:polyKernel}), Radial Basis Function (RBF) (eq. \ref{eq:rbfKernel}) and sigmoid (eq. \ref{eq:sigmoidKernel}).

\begin{equation} \label{eq:linKernel}
    k(a, b) = a \cdot b + c \textnormal{, where c = 1 for these experiments}
\end{equation}

\begin{equation} \label{eq:polyKernel}
    k(a, b) = (\gamma\langle{a, b}\rangle + r)^d
\end{equation}

For the polynomial kernel, the tested degrees were 2, 3, 4 and 5.
For all our experiments, the value of \emph{r}, which is also referred to as the \emph{0 coefficient} was 0.  
\begin{equation} \label{eq:rbfKernel}
    k(a, b) = \exp(-\gamma||a - b||^2)
\end{equation}

For the RBF kernel, we experimented with $\gamma \in \{10^{p_1}, 2^{p_2}\}$, where $p_1 \in \mathbb{N} \cap$[-2, 2] and $p_2 \in \{\pm15, \pm11, \pm9, \pm5, \pm3\}$. This hyperparameter controls the region of influence of a single example, where a lower value allows for a larger area (i.e., more influence).
\begin{equation} \label{eq:sigmoidKernel}
    k(a, b) = \tanh(\gamma\langle{a, b}\rangle + r)
\end{equation}

For the sigmoid kernel, $\gamma$ had the default value set for the scikit-learn implementation, equal to the $n * \sigma^{2}(X)){-1}$, where \emph{n} is the number of attributes and $\sigma^{2}(X)$ denotes the variance of X, while, as previously mentioned, \emph{r} was equal to 0. 

The other hyperparameter that was tuned as part of our experiments was the regularization parameter C, whose role is to trade between a simple decision and a low number of missclassifications. Thus, the lower its value, the smoother the decision surface; the higher its value, the fewer the incorrect predictions. The domain of C values explored was $\{10^{p_1}, 2^{p_2}\}$, where $p_1 \in \mathbb{N} \cap$[-2, 2] and $p_2 \in \{\pm15, \pm11, \pm9, \pm5, \pm3\}$, similar to \cite{2012JKAS...45...31C}. Nonetheless, it should be noted that however appealing a perfect classification is, this most commonly points to the issue of overfitting \citep[][]{doi:10.1021/ci0342472}, which manifests as the model having poor generalization skill, despite excellent training performance. We ensured the models learned to generalize well by examining the train and test errors at cross-validation and ensuring there were no high discrepancies between them.

Performance wise, out of all the SVM models applied on original data only, with balanced weights, the best results were obtained with the polynomial kernel, according to the F1 score, as shown in Table \ref{tab:svm_comparison_cv}.

\begin{table}[!ht]
    \caption{Performances of various SVM configurations on the entire dataset, with balanced weights, after 5-fold cross-validation. The 5 columns represent the following configurations: (1) - linear kernel; (2) - polynomial kernel, degree=2, C = 1, $\gamma$ = 1/(no. features $\cdot$ Var(X)); (3) - RBF kernel, C = $2^{-3}$, $\gamma$ = $2^{-15}$; (4) - RBF kernel, with uncertainty, C = $2^{-5}$, $\gamma$ = $2^{-11}$; (5) - linear kernel, with PCA, 5 components} 
    \begin{tabular*}{\textwidth}{l@{\extracolsep{\fill}}ccccc}
            \hline
             SVM Model & (1) & (2) & (3) & (4) & (5) \\
             \hline
             Recall & 90.16\% & 81.44\% & 71.69\% & 76.36\% & 89.51\%\\
             Precision & 8.92\% & 14.38\% & 14.50\% & 14.16\% & 8.73\%\\
             F1 score & 16.24\% & 24.43\% & 24.11\% & 23.89\% & 15.91\%\\
             Accuracy & 93.40\% & 96.40\% & 96.79\% & 96.55\% & 93.32\%\\
             Specificity & 93.42\% & 96.51\% & 96.51\% & 100.00\% & 93.35\%\\
             \hline
    \end{tabular*}
    \label{tab:svm_comparison_cv}
    \vspace{-4mm}
\end{table}

Despite the different performance values, some false positives and false negatives are common to all 3 models. There are 18 geoeffective CMEs that all these SVM models fail to identify, e.g., the event on 2005-06-09 14:36:05, actually labeled as "Very Poor Event" in the LASCO catalog, associated with a Dst = -106 nt, as well as 627 CMEs that the models incorrectly classify as geoeffective. Out of these 627 examples, 513 have the Central Position Angle = Angular Width = 360 (halo events) e.g. the event on 2013-12-07 07:36:05.

The experiment regarding the use of special weights for what we suspected could be potentially geoeffective events increased the recall value of the model having an RBF kernel with $\sim$5\%, accounting for a $<$1\% precision drop, while for the models with different kernels, the recall value dropped. The use of PCA lowered both the recall and the precision values, however, with $<$1\% each. These performances are also summarized in Table \ref{tab:svm_comparison_cv}, columns (4) and (5).

\subsection{Artificial Neural Networks}
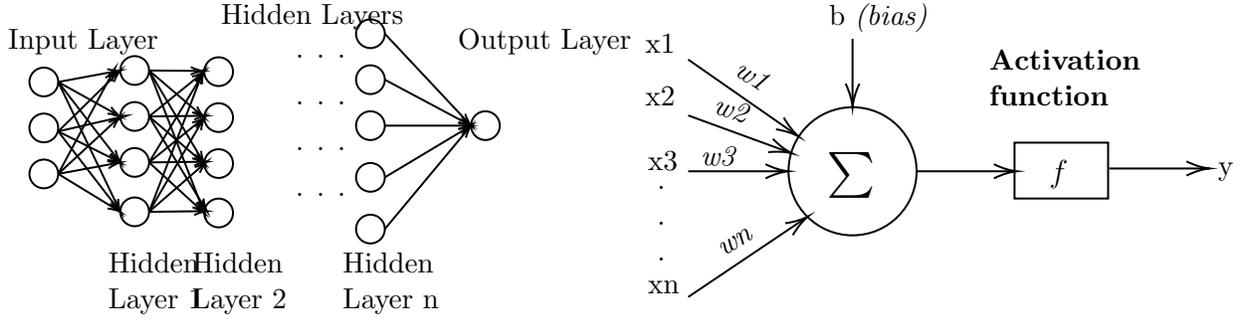
\begin{figure}[!ht]
    \centering
    \begin{minipage}{0.45\textwidth}
        \centering
        \input figures/anntikz.tex
    \end{minipage}
    \begin{minipage}{0.45\textwidth}
        \centering
        \tikzset{every picture/.style={line width=0.75pt}} 
\begin{tikzpicture}[x=0.75pt,y=0.75pt,yscale=-1.09,xscale=1.06, baseline=(XXXX.south) ]
\path (0,153);\path (293,0);\draw    ($(current bounding box.center)+(0,0.3em)$) node [anchor=south] (XXXX) {};
\draw   (74.11,82.71) .. controls (74.11,66.33) and (87.76,53.05) .. (104.59,53.05) .. controls (121.42,53.05) and (135.07,66.33) .. (135.07,82.71) .. controls (135.07,99.09) and (121.42,112.37) .. (104.59,112.37) .. controls (87.76,112.37) and (74.11,99.09) .. (74.11,82.71) -- cycle ;
\draw    (26.39,30.97) -- (77.42,65.55) ;
\draw [shift={(79.07,66.68)}, rotate = 214.13] [color={rgb, 255:red, 0; green, 0; blue, 0 }  ][line width=0.75]    (10.93,-3.29) .. controls (6.95,-1.4) and (3.31,-0.3) .. (0,0) .. controls (3.31,0.3) and (6.95,1.4) .. (10.93,3.29)   ;
\draw    (26.39,56.84) -- (74.01,73.77) ;
\draw [shift={(75.89,74.44)}, rotate = 199.57] [color={rgb, 255:red, 0; green, 0; blue, 0 }  ][line width=0.75]    (10.93,-3.29) .. controls (6.95,-1.4) and (3.31,-0.3) .. (0,0) .. controls (3.31,0.3) and (6.95,1.4) .. (10.93,3.29)   ;
\draw    (26.39,82.71) -- (72.11,82.71) ;
\draw [shift={(74.11,82.71)}, rotate = 180] [color={rgb, 255:red, 0; green, 0; blue, 0 }  ][line width=0.75]    (10.93,-3.29) .. controls (6.95,-1.4) and (3.31,-0.3) .. (0,0) .. controls (3.31,0.3) and (6.95,1.4) .. (10.93,3.29)   ;
\draw    (26.39,140.92) -- (81.86,104.64) ;
\draw [shift={(83.53,103.54)}, rotate = 506.81] [color={rgb, 255:red, 0; green, 0; blue, 0 }  ][line width=0.75]    (10.93,-3.29) .. controls (6.95,-1.4) and (3.31,-0.3) .. (0,0) .. controls (3.31,0.3) and (6.95,1.4) .. (10.93,3.29)   ;
\draw    (135.07,82.71) -- (180.79,82.71) ;
\draw [shift={(182.79,82.71)}, rotate = 180] [color={rgb, 255:red, 0; green, 0; blue, 0 }  ][line width=0.75]    (10.93,-3.29) .. controls (6.95,-1.4) and (3.31,-0.3) .. (0,0) .. controls (3.31,0.3) and (6.95,1.4) .. (10.93,3.29)   ;
\draw   (181.65,69.45) -- (226.19,69.45) -- (226.19,95.32) -- (181.65,95.32) -- cycle ;
\draw    (226.7,81.41) -- (272.42,81.41) ;
\draw [shift={(274.42,81.41)}, rotate = 180] [color={rgb, 255:red, 0; green, 0; blue, 0 }  ][line width=0.75]    (10.93,-3.29) .. controls (6.95,-1.4) and (3.31,-0.3) .. (0,0) .. controls (3.31,0.3) and (6.95,1.4) .. (10.93,3.29)   ;
\draw    (104.59,21.36) -- (104.59,51.05) ;
\draw [shift={(104.59,53.05)}, rotate = 270] [color={rgb, 255:red, 0; green, 0; blue, 0 }  ][line width=0.75]    (10.93,-3.29) .. controls (6.95,-1.4) and (3.31,-0.3) .. (0,0) .. controls (3.31,0.3) and (6.95,1.4) .. (10.93,3.29)   ;
\draw (5.94,132.21) node [anchor=north west][inner sep=0.75pt]   [align=left] {xn};
\draw (4.66,17.73) node [anchor=north west][inner sep=0.75pt]   [align=left] {x1};
\draw (4.66,42.31) node [anchor=north west][inner sep=0.75pt]   [align=left] {x2};
\draw (5.94,72.06) node [anchor=north west][inner sep=0.75pt]   [align=left] {x3};
\draw (10.3,88.16) node [anchor=north west][inner sep=0.75pt]  [font=\small] [align=left] {.\\.\\.};
\draw (52.24,28.21) node [anchor=north west][inner sep=0.75pt]  [rotate=-34.06] [align=left] {\textit{w1}};
\draw (39.65,45.67) node [anchor=north west][inner sep=0.75pt]  [rotate=-18.5] [align=left] {\textit{w2}};
\draw (31.15,69.63) node [anchor=north west][inner sep=0.75pt]  [rotate=-1.73] [align=left] {\textit{w3}};
\draw (38.09,116.58) node [anchor=north west][inner sep=0.75pt]  [rotate=-328.78] [align=left] {\textit{wn}};
\draw (277.18,77.12) node [anchor=north west][inner sep=0.75pt]   [align=left] {y};
\draw (196.83,76.26) node [anchor=north west][inner sep=0.75pt]  [font=\normalsize]  {$f$};
\draw (168.28,25.01) node [anchor=north west][inner sep=0.75pt]   [align=left] {\textbf{Activation}\\\textbf{function}};
\draw (92.02,3.5) node [anchor=north west][inner sep=0.75pt]   [align=left] {b \textit{(bias)}};
\draw (91,72.7) node [anchor=north west][inner sep=0.75pt]  [font=\Large]  {$\sum $};
\end{tikzpicture}
    \end{minipage}
    \caption{Left: the architecture of a feed-forward artificial neural network. Right: an artificial neural network neuron}
    \label{fig:ann}
\end{figure}

Feed-forward, fully connected artificial neural networks \citep[][]{Rosenblatt1958ThePA, 1986} are computational systems that map inputs to one or more outputs, having the following layered structure (fig. \ref{fig:ann} left), where each layer is fully connected to the previous one:

\begin{itemize}
    \setlength\itemsep{-0.5em}
    \item 1 input layer, representing the input variables and the bias
    \item 1 or more hidden layers, each with a previously defined number of neurons
    \item 1 output layer, representing the prediction
\end{itemize}

The number of both the layers and the neurons are hyperparameters that will be tuned.

The working principle of such a network is based on the neurons' way of working, which is similar to a linear regression, as shown in figure \ref{fig:ann} right. However, the complexity of the model is increased with the addition of the activation function.

Therefore, the output of the \emph{i$^{th}$} neuron in the hidden layer \emph{l} is expressed as:
\begin{equation}
    \label{neuronOutput}
    y_i^{[l]} = \varphi(W_i^{[l]T}X + b_i^{[l]})
\end{equation}

where $\varphi$ is the activation function and X are the outputs of all the neurons from layer \emph{l - 1}.

Neural networks are characterized by multiple hyperparameters, which will be discussed in the following paragraph. In terms of activation function, the sigmoid was used for the last layer, since the result is mapped to a number between 0 and 1, which can be interpreted as a probability for binary classification. For all other neurons, the Rectified Linear Unit \citep[ReLU,][]{inproceedings} (eq. \ref{eq:relu}) was chosen, being an adequate common activation function \citep[][]{nwankpa2018activation}.
\begin{equation}
    \label{eq:relu}
    ReLU(x) = max(0, x)
\end{equation}

The optimization algorithm used was Adam \citep[][]{kingma2017adam}, which has an adaptive learning rate whose default initial value is 1e-3 (for the TensorFlow and Keras implementations). Although it is considered that this algorithm needs little, if any, manual learning rate adjustments, we also tested the following values: 1e-1, 1e-2, 5e-2, 5e-3. The other specific parameters were unmodified. 
The loss function of choice was binary cross-entropy. The best batch size proved to be 32, out of the set of values we experimented with, i.e., 16, 32, 64, 128. The optimal number of epochs varied based on the number of neurons and learning rate. We employed the early stopping technique \citep[][]{1998} for finding the epoch when the loss value no longer decreases.
We trained and evaluated networks with 1 and 2 hidden layers, each with up to 10 neurons. Values higher than this proved to lead to either overfitting or no performance improvements, case in which less complex models that are easier to interpret are favoured. For these hyperparameter tuning experiments, the KerasTuner \citep[][]{omalley2019kerastuner} framework was used.  

\begin{table}[!ht]
\centering
    \caption{Performances of ANN models for various experiments, on the entire dataset, after 5-fold cross-validation. The headers represent the following 1 hidden layered architectures:\\
    (1) - 3 neurons, learning rate = 1e-3, batch size = 32, 25 epochs, balanced weights;\\
    (2) - 7 neurons, learning rate = 1e-3, batch size = 32, 23 epochs, balanced weights;\\
    (3) - 9 neurons, learning rate = 1e-3, batch size = 32, 30 epochs, with sample weights for uncertainty;\\
    (4) - 3 neurons, learning rate = 1e-3, batch size = 32, 23 epochs, after using PCA with 5 components, balanced weights;}
    \begin{tabular*}{\textwidth}{l@{\extracolsep{\fill}}cccc}
            \hline
             ANN Model & (1) & (2) & (3) & (4)\\
             \hline
             Recall & 92.40\% & 91.11\% & 90.34\% & 89.35\%\\
             Precision & 7.48\% & 7.82\% & 8.13\% & 5.34\%\\
             F1 score & 13.84\% & 14.41\% & 14.81\% & 10.05\%\\
             Accuracy & 91.88\% & 92.37\% & 92.64\% & 88.03\%\\
             Specificity & 92.92\% & 91.86\% & 92.52\% & 89.25\%\\
             \hline
    \end{tabular*}
    \label{tab:ann_cv}
    \vspace{-4mm}
\end{table}


What stood out for the ANN experiments was the similarity between the results from various experiments and architectures (including an additional hidden layer). A possible explanation for this may be the lack of essential data for distinguishing between the CMEs that are geoeffective and those that are not (e.g., interplanetary parameters) that such a model type could effectively use for better predictions.  

\subsection{Ensemble models}
\label{sec:ensemble}
Ensemble models \citep[][]{book} harness the power of two or more different models combined, in an attempt to achieve better prediction results than the individual ones. The results yielded by this approach can typically be interpreted as more confident ones, being obtained by combining the prediction capabilities of more than one metaphorical expert (i.e., model).  
For this research, a stacked classifier was employed. Stacked classifiers use the predictions of the models aggregated in the ensemble as inputs for an additional meta-learner. This learner is trained to find an optimal manner to combine the previous prediction in order to get a final label as accurate as possible. 

We decided to build an ensemble model with the aim of improving the precision, without significantly decreasing the recall value. Therefore, naturally, the ensemble needed to include both a model with high precision and one with high sensitivity. We decided on using the KNN model (with uniform weights, 7 neighbours and Manhattan distance), given that it has the highest precision. Additionally, a logistic regression model with balanced class weights and 0.9 threshold was used, given both its predictive performances and simplicity (e.g., compared to the more complex neural network). For the meta-learner, we chose a logistic regression model, which is also the scikit-learn default option for such classifiers, with the same class cutoff. 

\begin{figure}[!ht]
    \centering
    \caption{The predictions of the ensemble model, on the test set (20\% of the data), colored by their correctness}
    \label{fig:UMAPbest}
    \includegraphics[width=\textwidth]{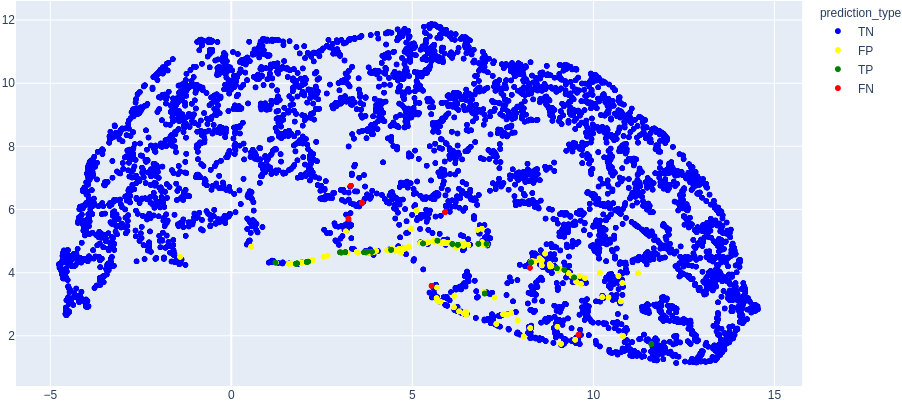}
\end{figure}

The recall values obtained for all models employed in this study generally varied between 70\% and 91\%, with precision values roughly fluctuating between 7\% and 16\%, with the exception of the KNN models, whose precision could reach 67.9\%, for a recall of 4\%. As previously mentioned, while a recall value as high as possible is desirable, we decided to choose the best model according to the highest F1 score value, and subsequently, the best precision value among models with a recall value of at least 75\%. Therefore, we consider this stacked classifier to be the best performing model. In order to obtain a visual representation of the correctness of the predictions, the dataset was randomly shuffled and split into 2 subsets, while preserving the 0-1 label ratio of the original dataset for all the subsets. 20\% of the data (4881 samples, out of which 33 positive ones) was set aside for final tests, while the remaining 80\% account for the \emph{traindev} (used for training and development) data. The model was trained on the traindev data and tested on the test set. The resulting 4 types of binary prediction categories, as summarized in Sec. \ref{sec:methods}, are showcased in Figure \ref{fig:UMAPbest}.

This model correctly classifies 4719 out of 4849 negative samples, represented by the blue dots in Figure \ref{fig:UMAPbest}. The 7 false negatives (red dots) are presented in Table \ref{tab:ensemble_fn}. 
However, out of these 7 false negative events, 5 are questionable events, having either poor/very poor CME data or an uncertain CME association.

\begin{table}[!ht]
\centering
    \caption{Example false negatives predicted by the ensemble model on the test set}
    \begin{tabular*}{\textwidth}{c@{\extracolsep{\fill}}cccccc}
            \hline
             CME & CPA & AW & LS & Acc & FI & Dst\\
             \hline
             1997-10-06 15:28 & 139 & 174 & 293 & 15.9 & 0.08 & -130\\
             2000-05-13 12:26 & 166 & 182 & 666 & 10.8 & 2.09 & -92\\
             2001-09-20 19:31 & 306 & 207 & 446 & -3.4 & 3.08 & -73\\
             2001-09-27 04:54 & 226 & 182 & 509 & 36.6 & 5.25 & -66\\
             2004-07-22 07:31 & 66 & 151 & 700 & 4.8 & 13.59 & -136\\ 
             2005-06-09 14:36 & 260 & 151 & 377 & -3.7 & 0.56 & -106\\
             2011-10-02 02:00 & 167 & 103 & 259 & 0.6 & 7.85 & -43\\
             \hline
    \end{tabular*}
    \label{tab:ensemble_fn}
    \vspace{-4mm}
\end{table}

In addition to this, the model predicted 130 false alarms, out of which 92 were full halos, having CPA = AW = 360. Analyzing Figure \ref{fig:UMAPbest} above, together with Figure \ref{fig:UMAP} left and right, while bearing in mind that the first one depicts only 20\% of the data (i.e., the test set), whereas the other two are representations of the entire dataset, the pattern followed by the samples incorrectly classified as geoffective can be easily observed. It is, therefore, more apparent how the similarities between some geoeffective samples and others pose a substantial challenge for the models. 

The stacked classifier qualifies as our best model, given its high recall and F1 score values, as noted in Table \ref{tab:ensemble_cv} (the second and third columns). This model has been employed for additional experiments, described below, that are summarized in the fourth and fifth columns.

\begin{table}[h]
\centering
    \caption{Performance of the stacked classifier on the full set with 5-fold cross validation, test, and subsampled data sets}
    \begin{tabular}{lcccc}
            \hline
             Stacking classifier & 
             Entire dataset & 
             Test dataset & 
             \makecell[l]{Samples with\\AW$ \geqslant$ 60} & 
             \makecell[l]{Samples with\\AW $\geqslant$ 120}\\
             \hline
             Recall &  80.28\% & 78.99\% & ~~~~~78.59\% & ~~~~~73.59\% \\ 
             Precision &  15.19\% & 14.82\% & ~~~~~17.12\%  & ~~~~~17.61\%\\
             F1 score &  25.54\% & 24.96\% & ~~~~~27.95\% & ~~~~~28.40\%\\
             Accuracy &  96.67\% & 96.65\% & ~~~~~90.69\% & ~~~~~74.27\%\\
             Specificity & 96.79\% & 96.78\% & ~~~~~90.90\% & ~~~~~74.42\%\\
             \hline
    \end{tabular}
    \label{tab:ensemble_cv}
    \vspace{-4mm}
\end{table}

\newpage
\subsection{Addressing the class imbalance}
\label{sec:class_imbalance}
We reiterate that the class imbalance represents one of the main challenges of this task. In section \ref{sec:challenges}, we discussed minimizing its negative effects by using of class weights by undersampling (decreasing the number of samples from the majority class) and oversampling (increasing the number of samples from the minority class.

We perform two additional experiments to attempt improving the prediction metrics: firstly, we attempted to artificially increase the number of geoeffective events, using SMOTE. The implementation used was provided by \emph{imbalanced-learn} library \citep[][]{JMLR:v18:16-365}.  

This technique relies on the principles of K-Nearest Neighbors (as detailed in section \ref{sec:knn}). A sample is considered to be a point in an n-dimensional space, where n is equal to the number of features. In order to generate a new sample, an existing point from the minority class is chosen randomly, together with k of the most similar points from the same class (i.e. nearest neighbors). Out of the k neighbors, one is randomly selected and a new point is generated, arbitrarily, on the line determined by the original point and its neighbor. Thus, a new sample has been created, its features' values being the coordinates of the newly generated point.

The sampling strategy was set to 0.1, meaning that new minority samples were created so as to obtain a 1-10 label ratio (i.e., 1 out of 10 samples would be geoeffective). In order to further reduce the imbalance to a 1-5 ratio with a reasonable amount of artificial data only, random samples from the majority class were eliminated (random undersampling with 0.2 sampling strategy). 

For this experiment, the traindev set was further split into 2 stratified subsets, further called train (80\%) and dev (20\%). SMOTE was only applied on the train set. The newly obtained set was then concateneted with the dev set, containing original data only, and shuffled. The resulting set was used for cross-validated experiments. We tested the models both on the entire dataset, containing all original and artificially generated samples, through 5-fold cross validation and on the test set containing original data only. The results are summarized in Table \ref{tab:smote_test}.
\begin{table}[ht!]
    \centering
    \caption{Performances of the models trained on the SMOTE augmented dataset, on the test set}
    \begin{tabular*}{\textwidth}{l@{\extracolsep{\fill}}cccc}
        \hline
         Model & \makecell[l]{Linear regression\\threshold = 0.9\\balanced weights} & \makecell[l]{KNN, 3 neighbours\\Manhattan distance\\distance-based weights} & \makecell[l]{SVM\\Linear kernel\\balanced weights} & ANN \\
         \hline
         Recall & 75.00\% & 43.75\% & 93.75\% & 71.87\%\\
         Precision & 17.77\% & 20.00\% & 10.20\% & 2.24\%\\
         F1 score &  28.74\% & 27.45\% & 18.40\% & 4.35\%\\
         Accuracy &  97.56\% & 98.48\% & 94.55\% & 79.28\%\\
         Specificity &  97.71\% & 98.84\% & 94.55\% & 79.33\%\\
         \hline
    \end{tabular*}
    \label{tab:smote_test}
    \vspace{-4mm}
\end{table}

Both the linear regression model and the KNN models exceed all F1 scores obtained using original data only. 

While we were able to build KNN and SVM models with higher precision values trained on original data only, their recall values in Table \ref{tab:smote_test} are higher than any obtained before. The SVM model, for example, only misses 2 geoeffective events, out of the 32 present in the test set: the events on 2005-06-09 14:36, 2011-10-02 02:00 (Dst values for these events being -43, -106 nT, respectively).

For the ANN we used the same architecture as model 1 (i.e., 1 hidden layer, 3 neurons, learning rate = 1e-3, batch size = 32, 25 epochs, balanced weights). This was the only model with lower performances. Nevertheless, it is important to bear in mind that the additional samples could showcase some particularities, since they were artificially generated, as well as the fact that one substantial disadvantage of this method is the lack of consideration regarding the positioning of the newly generated points in relation to other existent ones, from a different class. This means that any new point could have the exact same attribute values as a point from a different class, yet still have a different label. Additionally, the presence of this issue is all the more justified given that there are very few geoeffective examples to begin with and that, apart from the use of SMOTE, the issue of class overlap is, as previously stated, a challenge for this project in itself.  
 
Whilst we acknowledge the shortcomings of this technique and do not rule out the influence of other factors, these results indicating that having more positive samples to train on could improve the performances.

To pursue this, we present the second experiment, where we eliminate small-width CMEs, which are shown in the literature to be less likely to be heading towards the Earth \citep[][]{schw06,zhang2007,Chi2016}. We set two threshold levels, i. selecting CMEs having the AW $\geqslant$ 60, limiting our data set to 7373 samples, where the ratio of geoeffective samples raised from 0.8\% to 2.27\%. ii. selecting CMEs having AW $\geqslant$ 120, as analogous to \cite{2012JKAS...45...31C}), resulting in a subset of 2360 samples, out of which 6.90\% are geoeffective. We trained and tested the performance of the stacked classifier on these two subsampled data sets. While the precision and F1 score values increased (at the expense of the recall value), their improvement was not significant, as shown in Table \ref{tab:ensemble_cv}. 

As mentioned in section \ref{sec:lin_reg}, the choice of the threshold used for delimiting the two classes can influence the performance, as can be seen in Table \ref{tab:linreg_thresholds}. The results in Table \ref{tab:ensemble_cv} were obtained using 0.7 and 0.5, respectively, as thresholds for the logistic regressors used as part of the ensembles. These values were chosen empirically.

\begin{figure}[!h]
    \centering
    \begin{minipage}[b]{0.45\textwidth}
        \centering
        \includegraphics[height=0.17\textheight]{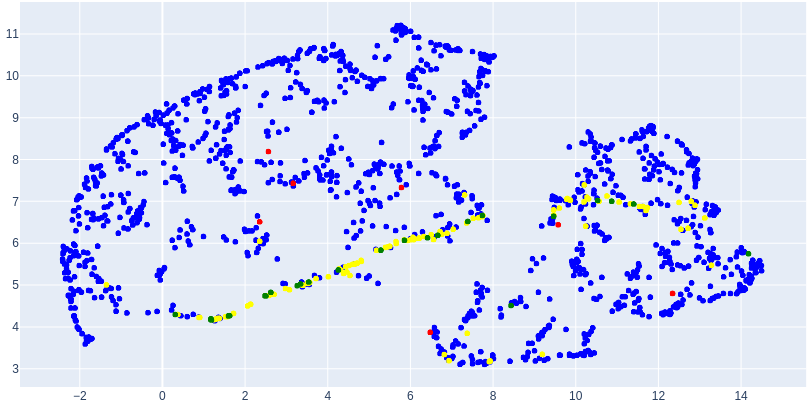}
    \end{minipage}
    \begin{minipage}[b]{0.45\textwidth}
        \centering
        \includegraphics[height=0.17\textheight]{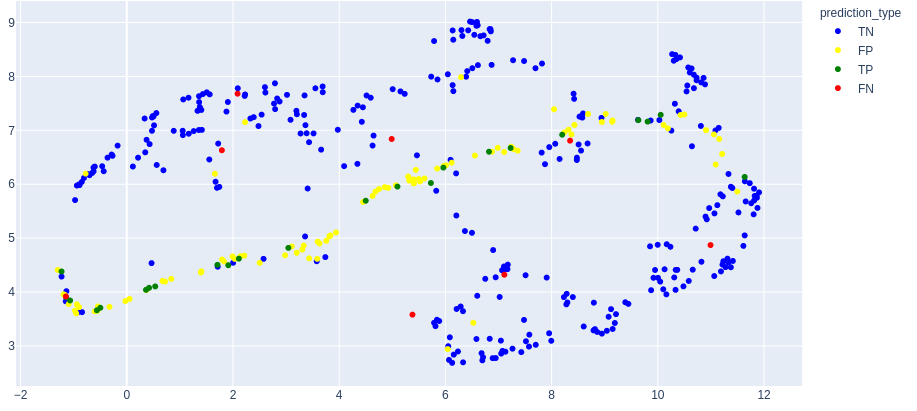}
    \end{minipage}
    \caption{Left: Predictions on the 1st subset (AW $\geqslant$ 60). Right: Predictions on the 2nd subset (AW $\geqslant$ 120)}
    \label{fig:UMAPsubsampled}
\end{figure}

The model still yielded 106 false alarms (130 for the full set), out of which 91 (92 for the full set) were full halos on the first subset, and 95 false alarms, out of which 85 full halos, on the second subset respectively, as seen in Figure \ref{fig:UMAPsubsampled} and Table \ref{tab:ensemble_cv}. This denotes that the inconspicuous differences between full halo CMEs that are geoffective and those that are not remained an issue even for less imbalanced sets. In conclusion, we could not clearly establish that the low number of geoeffective samples is the most influential factor that negatively impacts the model's prediction capabilities.

\newpage
\section{Conclusions and discussion}
\label{sec:conclusion}
The purpose of this study was to explore the usability of various ML methods for computing the probability that a given CME will be associated with a GS. For this investigation, we also assessed and addressed a variety of context specific issues, such as class imbalance and class overlap. We consider that, in the case of geoeffectiveness prediction, the sensitivity is the most important performance indicator, because of the high cost associated with a false negative in the eventuality of a powerful storm. Therefore, we believe that the recall values we obtained (e.g., 80.28\% for the stacked classifier) prove the practical utility of our models, given that only solar parameters have been used, which ensures extended warning times.

Apart from the numerical values of the performance metrics, a series of significant observations should be noted. One of the most important remarks emerged from this study is the association between the majority of the false alarms reported by our models with full halo CMEs. This is in agreement with the current beliefs that these ejections are the ones having the highest probability of being geoeffective, as well as being associated with the most powerful storms \citep[][]{2007JGRA..112.6112G}. Therefore, while the precision values are low, they are justified by the high rate of geomagnetic storms caused by full halo and/or fast CMEs (e.g. 113 out of 172 geoeffective CMEs had an angular width of 360; 537 CMEs having an angular width of 360 were not geoeffective). One notable remark is that halo CMEs are observed equally well originating on both the frontside and the backside of the Sun. While backside CMEs do not reach Earth and, hence, are not geoeffective, the observations and catalogs used for this study do not discern between the two classes. Therefore, we did not include such separation in the model setups. An additional consideration is that threshold adjustments were the most effective technique for increasing precision values. Furthermore, using the original values of the independent features led to better results than using the values obtained after lowering the number of dimensions by using PCA, indicating the original CME attributes used are appropriate inputs for such studies. Lastly, most models trained on SMOTE augmented data exceeded the average performance values obtained with real data only, on the test set. 

The present study adds a number of original contributions to the growing body of research on the application of ML in the field of heliophysics. To the best of our knowledge, no previous studies focused on this topic created artificial data samples using SMOTE, introduced mechanisms for dealing with potentially geoeffective events or employed UMAP for data visualization.

Our experiments use a considerably more extensive dataset than the ones used by \cite{srivastava2005} (64 geoeffective CMEs) and \cite{10.3389/fspas.2021.672203} (2796 CMEs, 32 geoeffective ones), who employ logistic regression models for predictions. 

Our results are also comparable with those obtained by \cite{2012JKAS...45...31C}. We used a more extensive dataset, i.e., not limited to halo CMEs, and incorporated minor CMEs  with -50 nT $<$ Dst $\leqslant$ -30 nT. Their SVM model (RBF kernel, C = $2^{-5}$, $\gamma$ = $2^{-15}$) obtained 76\% sensitivity, together with 28\% precision, based on a narrower definition for geoeffectiveness (where a geoeffective event is defined as being associated with Dst $<$ -50 nT in that work), therefore not including weak storms. Most of our models, including the stacked classifier, compare favourably to this in terms of the proportion of geoeffective CMEs identified, while in turn obtaining a lower precision, which might be explained by our inclusion of both non-halo CMEs and weak storms in our dataset. 

Similarly, \cite{uwamahoro2012} only include GS with Dst $\leqslant$ -50 nT and halo CMEs in their study on using neural networks for predicting the geoeffectiveness. Their results (86\% sensitivity, 81.4\% precision) are obtained with a 3 layer network that takes interplanetary data as inputs, in addition to solar ones. While these values appear promising, the use of interplanetary parameters substantially limits the warning time to less than a day, given that this data is collected closer to Earth. This is the reason why the authors also emphasize the importance of prediction models that only use solar parameters, which are accessible earlier and can lead to up to 3-4 days of warning time. Such a model could be used for filtering out most CMEs with high chances of not being geoeffective, therefore creating warnings for potentially geoffective ones with days in advance. Their actual geoeffectiveness could, then, be estimated with more complex methods and higher precision when more ongoing data has been collected.   

\cite{DBLP:journals/remotesensing/FuZYFLM21} obtained a similar F1 score (27\%), with a lower accuracy value (75\%) using deep learning with only solar images as inputs. Although the main scope of their work is different (i.e., predicting the arrival time) and the methods cannot be directly compared, the difference between accuracy values is notable.  Our future goals consist of assessing the feasibility of augmenting our method with image based ML methods and results, with the hope of enhancing the results presented here.

We consider our results to be promising in terms of recall values, in spite of the limitations imposed by the nature of the data. This research shows that, using only the independent CME solar attributes available online, easily interpretable models such as logistic regressions can correctly identify over 75\% of the geoeffective CMEs. In fact, the increased complexity of the model (e.g., neural network) did not translate, in this case, into substantially increased prediction performances. Our best model is the stacking classifier with 2 estimators: a logistic regression with a 0.9 threshold for discerning between classes and a KNN model with uniform weights, 7 neighbours and Manhattan distance, tied together by a meta learner in the form of an additional logistic regression, with balanced weights. We considered this to be the best performing model, based on having the highest value for the recall and f1 score pair, thus having the best recall-precision balance (based on f1 score), while also identifying most of the geoeffective events (based on recall). This model correctly identified 80.28\% of the geoeffective events, with a precision of 15.19\%, based on 5-fold cross-validation on the entire dataset , as shown in Table \ref{tab:ensemble_cv}.

We believe the results obtained in this study might be enhanced by including comprehensive flare associations. While such associations were not included in this work, this could be a fruitful direction for further exploration. An example of such an association is the flare to CME association list of \citet{yashiroetal2006}. This, however, only includes flares of M1.0 class, which would severely limit our already small sample of geoeffective events. Thus, it may be significant to expand the \citet{yashiroetal2006} list to include smaller C, or even B flares that might be linked to geoeffective CMEs from filament eruptions. We believe a comprehensive set of associations covering the majority of the geoeffective events studied here would sort out additional uncertainty factors, such as the back-halo CMEs, and might significantly improve the model metrics described in this work.

Lastly, we mention that the underlying challenges associated with this task, namely the extreme class imbalance, the low number of variables available at the moment and the issue of class overlap, remain significant obstacles for this type of predictions. This exploratory study presents encouraging results using numerical data only. We are further exploring the extension of our research to include solar image data, as well as more complex learning frameworks.   


\begin{acknowledgments}
The authors thank Dr. Sarah Gibson for the initial review of this work. We are grateful for the suggestions offered by the anonymous reviewer, which have significantly enhanced this work. 
A.R.P. was funded by the High Altitude Observatory of the National Center for Atmospheric Research, facilities that are sponsored by the National Science Foundation under cooperative agreement No. 1852977.
We acknowledge the use of the SOHO/LASCO CME catalog. This CME catalog is generated and maintained at the CDAW Data Center by NASA and The Catholic University of America in cooperation with the Naval Research Laboratory. SOHO is a project of international cooperation between ESA and NASA. Flare Index Data used in this study were calculated by T.Atac and A.Ozguc from Bogazici University Kandilli Observatory, Istanbul, Turkey and made available through the NOAA National Geophysical Data Center (NGDC).
\end{acknowledgments}



\bibliography{bandr}{}
\bibliographystyle{aasjournal}

\end{document}